\documentstyle[prb,aps,epsf,eqsecnum,multicol,epsfig,psfrag]{revtex}

\newcommand{\beq}{\begin{equation}}
\newcommand{\enq}{\end{equation}}  
\begin{document}
\bibliographystyle{prsty}

\title{Plasticity in current-driven vortex lattices}

\author{Panayotis Benetatos}
\address{Hahn-Meitner-Institut, Abteilung Theoretische Physik (SF5), Glienicker Str. 100, D-14109, Berlin, Germany}
\author{M. Cristina Marchetti}
\address{Physics Department, Syracuse University, Syracuse, NY 13244, USA}
 
\date{\today}
 
\maketitle
  
\begin{abstract}
We present a theoretical analysis of recent experiments on
current-driven vortex dynamics in the Corbino disk geometry. 
This geometry introduces controlled spatial  
gradients in the driving force and allows the study of the onset of
plasticity and tearing in clean vortex lattices. 
We describe  plastic slip in terms of the stress-driven unbinding 
of dislocation pairs, which in turn contribute to the relaxation
of the shear, yielding a nonlinear response. 
The steady state density of free dislocations induced by the applied stress
is calculated as a function of the applied current and temperature.
A criterion for the onset of plasticity at a radial location $r$ in the disk yields a temperature-dependent
critical current that is in
qualitative agreement with experiments.

\end{abstract}

\begin{multicols}{2}

\section{Introduction}

In the mixed state of type-II superconductors the magnetic field is concentrated in
an array of flexible flux bundles that, much like ordinary matter,
can form crystalline, liquid and glassy phases.\cite{CrabtreeNelson,Blatter} 
In clean systems the vortex solid melts
into a flux liquid via a first order phase transition. 
\cite{CrabtreeNelson}
If the barriers to vortex line crossing are high, a rapidly cooled vortex 
liquid can bypass the crystal phase and get trapped in a 
metastable polymer-like glass phase,
much like ordinary window glass.\cite{nels}
The diversity of vortex structures
is further increased by pinning from material disorder, which leads to a
variety of novel glasses. 
\cite{FisherFisherHuse,NelsonVin} 

Of particular interest is the {\it dynamics} of the vortex array in the various phases and in the
proximity of a phase transition. In the liquid phase the vortex array flows yielding a linear resistivity.
In the presence of large scale spatial inhomogeneities, the liquid flow can
be highly nonlocal due to interactions and entanglement.
\cite{mcmdrnhyd90,HuseMajum93}
The correlation length controlling the spatial
nonlocality of the flow
grows with the liquid shear viscosity, which becomes large 
as the liquid freezes. At a continuous liquid-glass transition
this correlation length diverges with a universal critical exponent.\cite{mcmdrn00}
In general, probing velocity correlations in driven vortex arrays
yields information
on vortex dynamics within a given phase, as well as on the nature of the
phase transitions connecting the various phases.

As for ordinary matter, the shear rigidity of the vortex array 
can be probed by 
forcing the vortices to flow in confined geometries
obtained by engineering suitable artificial pinning structures,
as discussed for instance in Ref. \onlinecite{mcmdrn00}.
Large scale spatial inhomogeneities can also be introduced in the flow, even in the absence
of pinning, by applying a driving force with controlled spatial gradients,
as done recently 
by the Argonne group using the Corbino disk geometry 
sketched in Fig. 1.\cite{Argonne_Co1,Argonne,Argonne_Co2} 
In the Corbino disk  a radial driving current, $I$, yields a spatially
inhomogeneous Lorentz force that decreases as $\sim 1/r$, 
with $r$ the distance from the center of the disk.
For small applied currents, the local shear stresses are negligible and 
the vortex array moves as a rigid body.
Larger currents (or even vanishingly small currents
in a glassy solid) result in a
strong spatially inhomogeneous stress, $\sigma(r,I)$. The solid 
``breaks up'' in concentric annular regions flowing at different velocity
and the response is highly nonlinear.\cite{Argonne}
The voltage profiles measured experimentally by placing a series
of voltage taps in the radial direction (see Fig. 1) reflect the 
different dynamical correlation lengths in the fluid, plastic and elastic
regimes. In a given experiment, shear-induced plastic slip is observed to occur
at different values of the applied current in different regions of the sample.
This is a direct consequence of the controlled spatial inhomogeneity 
of the shear stress. 


\begin{figure}
\begin{center}
\label{Corbino_geometry}
\leavevmode
\hbox{%
\epsfxsize=2.6in
\epsffile{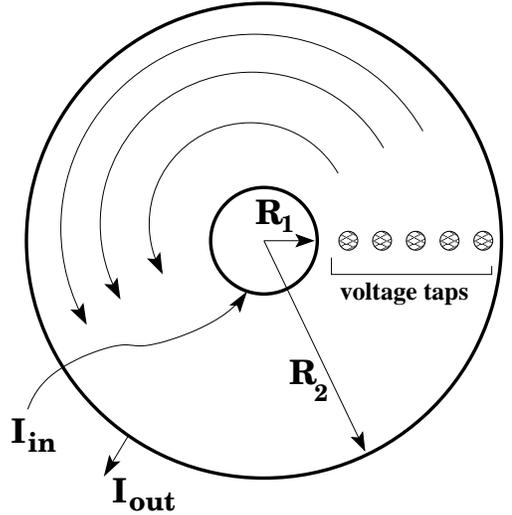}}
\end{center}
\caption{The Corbino disk geometry. The external magnetic field
is out of the page. The electrical current flows radially from the
inner circumference to the outer rim of the disk. 
The radial current density gives rise
to a Lorentz force which causes the vortices to moves in circular orbits
around the disk, without crossing the sample edges. }
\end{figure}


\noindent Slippage occurs first in the inner region of 
the disk where the Lorentz force is largest, and propagates to the outer regions
as the current is increased. 
In this regime of plastic response the dynamical correlation length 
controlling velocity inhomogeneities
can be identified with the separation between free dislocations
that are continuously created and annihilated by the strong local shear stresses.

In this paper, we present a theoretical analysis of these
experiments. A preliminary account of part of this
work was presented in Ref. \onlinecite{mcmHouston}.
We describe the onset of
plasticity in the driven vortex solid as a nonlinear effect 
due to the stress-induced proliferation of unbound dislocations.
The static shear stress from the applied current yields a Peach-Koehler 
force that pulls apart
neutral, bound dislocation pairs present in the lattice
in equilibrium.
As a result, a finite density, $n_f(\sigma,T)$, of
free dislocations is generated in the solid. Such unbound dislocations
can then 
contribute to
the relaxation of the  shear, resulting in a highly nonlinear
response to the applied current. This mechanism for nonlinear
shear relaxation has been studied before in the context of 
superfluid films 
\cite{AHNS80}, $2D$ solids \cite{BHZ,ZHN80}
and smectic liquid crystal films \cite{franosch}.
In all this cases, the response to a spatially inhomogeneous shear 
was considered.
Here we adapt this work to the case where the external shear is 
spatially inhomogeneous, but still slowly varying on the length scales of 
interest.

Although this paper deals specifically with the onset of 
plasticity in vortex lattices in the Corbino disk geometry,
the methods and results presented have a more general interest.
In particular, the Langevin and Fokker-Planck descriptions
of dislocation dynamics in a spatially-inhomogeneous stress field
can be used to describe nonlinear shear relaxation in other systems
and geometries. Our work is for instance  relevant
to the defect dynamics in two-dimensional colloidal crystals,
where individual point defects can be  created and manipulated 
with optical tweezers.\cite{pertsinidis1,pertsinidis2} 
Although only vacancies and interstitials have been 
considered so far, 2d colloids, where direct real space and time imaging of
defects is possible,  may be especially suitable 
for studying in details the dynamics of dislocations in the presence
of  external stresses.

We begin in Section II with a brief review of 
the Corbino disk geometry and of the 
qualitatively different response of a vortex liquid and a vortex solid to an 
applied radial current. 
In the vortex solid, static elastic deformations 
yield a local shear stress $\sigma(r,I)$, which,
for the case of free boundary conditions, decreases as $\sim 1/r^2$.
This stress can break bound dislocation pairs present in 
the solid in equilibrium, as shown in  sections III and IV.
For a finite external stress, the total interaction energy
of a neutral dislocation pair exhibits a saddle point with a finite barrier
at a value $x_c\sim1/\sigma(r)$ of the pair
separation. At zero temperature, the pair unbinds when the 
stress is large enough that $x_c(\sigma)\sim a_0$. At finite temperature
unbinding takes place when the pair escapes via thermal activation over the barrier. Using Langevin and Fokker-Planck descriptions of the dynamics
of bound pairs, we evaluate the rate of thermal escape and use
this to estimate the density of free dislocation,
$n_f(\sigma,T)$.  
The onset of plasticity is then identified with the proliferation of
unbound dislocations. 
For free boundary conditions (which are appropriate for the Argonne 
experiments), the applied stress is largest near the center
of the disk and,  for a fixed current $I$, dislocations
unbind first near the inner rim.
We can then evaluate the critical current $I_{\rm pl}(r,T)$
where the onset of plastic slip occurs as a function of temperature $T$ 
and of the radial distance
from the center,
with the result
\beq
\label{Iplastic}
I_{\rm pl}(r,T)=\frac{I_0(T)}{1+A/r^2}\;,
\enq
with $A=2R_1^2R_2^2\ln(R_2/R_1)/(R_2^2-R_1^2)$ a geometrical factor.
The characteristic current scale  
$I_0(T)= 4cc_{66}t/B_{0}\exp[-b/(T_M-T)^{1/2}]$ 
is the current where shear
induced proliferation of dislocations occurs across the entire sample,
with $T_M$ the 
Kosterlitz-Thouless melting temperature and $b$ a numerical 
constant.\cite{KT_note}
As observed in experiments, for fixed $T$, the current $I_{\rm pl}$ increases
away from the center of the disk. Conversely, for fixed $r$, it is strongly
temperature dependent and it decreases as the melting point 
$T_M$ is approached from below.

\section{The Corbino disk geometry}
The Corbino disk geometry was  used recently by the Argonne group
to introduce controlled spatial inhomogeneities in driven 
vortex arrays and study
the onset of plasticity in clean samples,
near the first order melting transition.\cite{Argonne}
The same  geometry has
been used before to minimize edge effects \cite{paltiel00} and also to study the 
Hall effect in type-I superconductors \cite{HallCorb}.

The Corbino disk is an annular
superconducting slab placed in a magnetic field parallel to its axis
(chosen here as the $\hat{\bf z}$ direction), as shown in Fig. \ref{Corbino_geometry}. 
A current $I$ is injected at the center of the disk and removed at the outer boundary,
creating a radial current density,
\beq
\label{curdens}
{\bf J}(r)=\frac{I}{2\pi r t}\;\hat{\bf r}\;,
\enq
where $r$ is the radial distance from the center and $t$ the thickness
of the disk. 
With the magnetic field applied along the disk axis, the Lorentz force density
on the vortices is azimuthal, with 
\beq
{\bf f}_L(r)=\frac{1}{c}B_0J(r)\bbox{\hat{\phi}}\;,
\enq
and drives the vortices to move in circular orbits around the disk,
without crossing the sample edges.
Here, $B_0=n_0 \phi_0$ is the average magnetic induction in the
sample and $\bbox{\hat{\phi}}=\hat{\bf z}\times\hat{\bf r}$ is the unit vector
in the tangential direction. In this geometry, the driving force
is spatially inhomogeneous and stronger near the center of the disk,
falling off as $\sim 1/r$. It naturally yields spatially inhomogeneous
distortions of the vortex array that can be probed by placing a series
of voltage taps in the radial direction.
As described below, the spatial dependence of the voltage profile in this
geometry is qualitatively different in the vortex liquid
and solid phases and provides a natural way to study the onset of
plasticity.

As described in Ref. \onlinecite{Argonne}, in the liquid state, 
vortex motion is governed by the hydrodynamic
equation for the vortex flow velocity ${\bf v}$, \cite{mcmdrnhyd90,mcmdrnPhysica91}
\beq
\label{veldiffeq}
-\gamma {\bf v} + \eta {\nabla}_{\perp}^2 {\bf v}=-{\bf f}_L\;,
\enq
where $\gamma(T, H)$ is the friction and $\eta(T, H)$ the shear
viscosity of the vortex liquid. The flow profile can be found by
solving Eq. (\ref{veldiffeq}) with suitable boundary conditions. \cite{mcmHouston}
The electric field induced by the vortex motion is then obtained from
\beq
{\bf E}=\frac{1}{c}{\bf B}\times {\bf v}\;.
\enq
If the radial width of the disk is much larger 
than the viscous penetration length,
$\delta=\sqrt{\eta/\gamma}$, the effect of the boundaries is negligible
and the velocity profile simply follows the force profile, with
${\bf v}={\bf f}_L/\gamma$, i.e., $v\sim 1/r$. 
The voltage drop between the $n$-th and the
$n+1$-th electrodes is then given by
\begin{equation}
\label{hydvolt}
V_{n, n+1}=\int_{R_n}^{R_{n+1}}E(r)dr=\frac{B_0^2 I}{\gamma c^2 2 \pi t}
\ln{\bigg(\frac{R_{n+1}}{R_n}\bigg)}\;.
\enq
where $R_n$ is the radial location of the $n$-th electrode.
This logarithmic scaling of the voltage characterizes fluid flow,
with neglible spatial inhomogeneities caused by viscous effects
at the boundaries. 
 
In contrast, a vortex lattice with a finite shear modulus rotates as 
a rigid body due to the azimuthal applied force,
provided the shear stresses due to the local force gradients 
are not too strong. 
This rigid body rotation yields a local velocity that scales
as $\sim r$, with

\beq
\label{elastvolt}
V_{n, n+1}=\int_{R_n}^{R_{n+1}}E(r)dr=\frac{B_0\omega}{2c}
          \big(R_{n+1}^2-R_n^2\big)\;,
\enq
where $\omega$ is the angular velocity of rotation of the vortex
lattice, given by 
\beq
\label{ang_omega}
\omega=\frac{B_0I}{\pi c\gamma(R_2-R_1)}\frac{1}{R_2^2+R_1^2}\;,
\enq
with $R_1$ and $R_2$ the inner and outer radii of the disk, respectively.
Above a certain value of the applied current, the local shear stresses
become strong enough to break the lattice bonds, or generate unbound
dislocations. In this regime, one obtains a plastic response
where
the lattice  ``breaks'' into two or more concentric
annular sections rotating at
different angular velocities and slipping past each other 
due to the continuous generation and recombination
of topological defects. Such a plastic response
is  evident in the experiments by the Argonne group.\cite{Argonne}

The spatial dependence of the Lorentz force makes it possible for all of 
the types of reponse mentioned above
to coexist at  a given value of the applied current. 
Near the inner radius of the disk the shear stress is very large and 
the vortex array flows like a liquid, with a logarithmic scaling of the voltage
drop with the contact positions.
In the middle of the disk the motion is plastic, with dislocations continuously
being created and annihilated. Near the outer radius
the shear stress is very small and the vortex array rotates as a 
rigid body.
Evidence for the coexistence of these behaviors was
obtained in Ref. \onlinecite{Argonne} by measuring the potential drops across 
successive pairs of
electrodes.

In this article, we describe the onset of
plasticity in the driven vortex solid as a nonlinear effect 
due to the stress-induced proliferation of unbound dislocations.
The spatially inhomogeneous Lorentz force gives rise to
elastic deformations of the vortex lattice described by the
solution of the equation 
\beq
\label{dynamics}
{\bf f}_{\rm el}+ {\bf f}_L(r)=0\;,
\enq
where
${\bf f}_{\rm el}$ is the elastic force density, given by,
\beq
\label{elforce}
f_{{\rm el},i}=\partial_j\sigma_{ij}\;,
\enq
with $\sigma_{ij}$ the stress tensor,
\beq
\label{stress_tensor}
\sigma_{ij}=2c_{66}u_{ij}+(c_{11}-c_{66})\delta_{ij}u_{ll}\;,
\enq
and 
$u_{ij}=(\partial_ju_i+\partial_iu_j)/2$ the symmetrized strain tensor.
Elastic deformations are described in terms of 
the  two-dimensional displacement field,  ${\bf u}$,
and $c_{66}$ and $c_{11}$ are the shear and
compressional moduli of
the vortex lattice, respectively.
Since dense vortex arrays are practically
incompressible  ($\;c_{11}>>c_{66}\;$), 
Eq. (\ref{dynamics})  becomes simply
\beq
\label{eleq_incomp}
c_{66}{\nabla}_{\perp}^2{\bf u}=- {\bf f}_L(r)\;.
\enq
The only nonvanishing component of the displacement field is in
the tangential direction, 
${\bf u}({\bf r})=u_\phi(r)\;\bbox{\hat{\phi}}\;$ and yields
a stress $\sigma_{r\phi}=2c_{66}u_{r\phi}\;$.
For free boundary conditions \cite{bc_note} 
at the inner and outer circumference of the disk,
$[\partial_r {\bf u}]_{r=R_1}=[\partial_r {\bf u}]_{r=R_2}=0\;$, we obtain
\beq
\label{stress}
\sigma_{r\phi}(r)=-\frac{IB_{0}}{4c\pi t}\bigg[1+
\frac{1}{r^2}\ln{(R_2/R_1)}\frac{2R_1^2R_2^2}{R_2^2-R_1^2}\bigg]\;.
\enq
The spatially inhomogeneous stress given by Eq. (\ref{stress}) 
decreases as $1/r^2$ and
gives rise to a Peach-Koehler
force that can unbind dislocations from bound pairs, yielding
free dislocations that in turn contribute to
the relaxation of the applied shear.

\section{Dynamics of neutral dislocation pairs}

Both edge and screw dislocations can occur in the vortex lattice.
The geometry and properties of such dislocation lines has been discussed for instance in Ref. \onlinecite{mcmdis90}. For simplicity here
we assume that the disk is sufficiently thin that thermally-induced
vortex wandering in the direction transverse to the applied field is
negligible and the vortices are essentially rigid rods.\cite{CarmenM}
In this limit, only edge dislocations can occur in the Abrikosov lattice.
The dislocation lines are aligned with the $z$ direction and their 
Burgers vectors lie in the $xy$ plane. The geometry and properties of such
rigid dislocation lines are the same as those of point dislocations in
two-dimensional lattices. In particular, the bare interaction energy of
an isolated pair of dislocations of Burgers vectors ${\bf b}^{(1)}$
and ${\bf b}^{(2)}$, located at ${\bf r}_1$ and ${\bf r}_2$,
is given by,
\end{multicols}
\begin{eqnarray}
\label{bare_int12}
U_0^B({\bf r}_1-{\bf r}_2)= & & \frac{K_0}{4\pi}\bigg\{
 {\bf b}^{(1)}\cdot{\bf b}^{(2)}\ln\bigg({|{\bf r}_1-{\bf r}_2|\over a_0}\bigg)
  -\frac{{\bf b}^{(1)}\cdot({\bf r}_1-{\bf r}_2)~
           {\bf b}^{(2)}\cdot({\bf r}_1-{\bf r}_2)}
      {|{\bf r}_1-{\bf r}_2|^2}\bigg\}\\\nonumber
 & &         +\frac{E_c t}{a_0^2}\big[|{\bf b}^{(1)}|^2+|{\bf b}^{(2)}|^2\big],
\end{eqnarray}
\begin{multicols}{2}
\noindent where 
$E_c\sim c_{66}a_0^2$ is the core energy per unit length of an
edge dislocation and the coupling constant $K_0$ is given by
\beq
\label{Kbare}
K_0=4c_{66}(c_{11}-c_{66})t/c_{11}\approx 4c_{66}t\;.
\enq
The last approximate equality in Eq. (\ref{Kbare}) holds for incompressible lattices. The (dimensionful) Burgers vector ${\bf b}$ is defined 
as the jump in the displacement ${\bf u}$ upon integration around a closed
contour,
\beq
\label{Burgers}
\oint d{\bf u}={\bf b}\;.
\enq
We are interested here in the nucleation of free dislocations
from bound pairs.  The interaction energy of a neutral pair, consisting of two 
dislocations of opposite Burgers vectors,
${\bf b}^{(1)}=-{\bf b}^{(2)}=-{\bf b}$,  is
\beq
\label{bare_int}
U_0^B(\bbox{\rho})=\frac{K_0a_0^2}{4\pi}\bigg[
                \ln\bigg(\frac{|\bbox{\rho}|}{a_0}\bigg)
               -\cos^2\theta\bigg] +2E_c t\;,
\enq
where
$\bbox{\rho}={\bf r}_1-{\bf r}_2$ and
$\theta$ is the angle between $\bbox{\rho}$ and ${\bf b}$.
The quantity $2E_ct$ represents the energy to create a pair of 
straight edge dislocations
of length $t$ 
at a distance $a_0$ relative to a dislocation-free system. It plays the role 
of a chemical potential for dislocation pairs.
The strictly two-dimensional limit of point vortices (as opposed
to the case of rigid vortex lines considered here) is recovered by 
the replacement 
$c_{66}t\rightarrow c_{66}^{2d}$, where $c_{66}$ and $c_{66}^{2d}$
are the shear moduli of a three-dimensional and of a two-dimensional lattice,
respectively.

In general, many dislocation pairs will be present in the lattice.
The interaction of a given pair is then renormalized by a screening
cloud of other pairs. Following Kosterlitz and Thouless,\cite{KT_note} 
this effect can be described 
in terms of a scale-dependent dielectric constant, $\epsilon(\rho)$.
Neglecting the angular part of the dislocation interaction,
which is a marginal perturbation at large length scales
\cite{nelson78},  the effective interaction of a neutral pair is given by
\beq
\label{eff_int}
U_0(\rho)=\frac{K_0a_0^2}{4\pi}\int_{a_0}^\rho
   \frac{d\rho'}{\epsilon(\rho')\rho'}+2E_c t\;.
\enq
At equilibrium, in the absence of external stresses, the probability
per unit area $\Gamma_0(\rho)$
of finding a pair of dislocation of separation $\rho$
is given by 
\beq
\label{Gamma_eq}
\Gamma_0(\rho)=\frac{1}{a_0^4}\exp\Big[-U_0(\rho)/k_BT\Big]\;.
\enq
and remains very small away from the Kosterlitz-Thouless
melting transition, $T_M=K_0a_0^2/(16\pi k_B)$.
Following Ref. \onlinecite{franosch},
the density of pairs on scales
$\rho$ is equivalently defined by the equation,
\beq
\label{Gamma_rec}
\frac{d\Gamma_0}{d\rho}=-\frac{\overline{K}_0}{\epsilon(\rho)\rho}
   \Gamma_0(\rho)\;,
\enq
with
\beq
\overline{K}_0=\frac{K_0a_0^2}{4\pi k_BT}
\enq
a dimensionless coupling constant.
Equation (\ref{Gamma_rec})  is naturally rewritten in terms of a scale dependent stiffness,
$\overline{K}(\rho)=\overline{K}_0/\epsilon(\rho)$, and a dislocation 
fugacity, $y(\rho)$, with $[y(\rho)]^2=\rho^4\Gamma_0(\rho)$.\cite{y_note}
The bare fugacity is simply $y_0=y(a_0)=\exp\big[-E_ct/k_BT\big]$.
One can then see immediately that Eq. (\ref{Gamma_rec}) is nothing but the
first of the Kosterlitz-Thouless recursion relations, usually written in
terms of the fugacity as
\beq
\label{fugacity_rec}
\frac{dy}{d\ln\rho}=\Big[2-\frac{\overline{K}(\rho)}{2}\Big]y\;.
\enq
The scale-dependence of the stiffness arises from the polarization of 
dislocation pairs and is governed by the second KT recursion relation,
\beq
\label{K_rec}
\frac{d\overline{K}^{-1}}{d\ln\rho}=2\pi^2y^2\;.
\enq
It is well known that the KT flow equations yield a low-temperature phase
with finite long-wavelength dielectric constant, separated by a
 high-temperature phase where $\epsilon(\rho)\rightarrow\infty$ at large
scales and unbound dislocations proliferate. The melting
occurs at $\overline{K}=4$. \cite{KT_note}

To discuss the nonequilibrium nucleation of free dislocations from bound
pairs under the action of an  applied stress, we need to study the dynamics
of neutral dislocation pairs as they move under the action of their mutual interaction
and of the Peach-Koehler force due to the external stress field.
The position ${\bf r}_\nu$ of the Burgers
vector ${\bf b}^{(\nu)}$ is assumed to obey a Langevin equation
of the form,\cite{peierls_note}
\begin{eqnarray}
\label{Langevin}
\frac{dr_{\nu i}}{dt}=\mu^{(\nu)}_{ij}\bigg[ & &
           -\sum_{\mu\not=\nu}\frac{\partial U_0({\bf r}_\nu-{\bf r}_\mu)}
        {\partial r_{\nu j}}+F^{\rm PK(\nu)}_{j}({\bf r}_\nu)\bigg]\\\nonumber
      & &  +\eta^{(\nu)}_{i}(t)\;,
\end{eqnarray}
where Latin indices denote Cartesian components, 
$\mu^{(\nu)}_{ij}$ is the mobility tensor, and $\eta^{(\nu)}_i$ 
is a random white noise, with
\beq
\langle \eta^{(\nu)}_i (t) \eta^{(\mu)}_j(t')\rangle=
         2k_B T\mu^{(\nu)}_{ij}\delta_{\mu\nu} \delta(t-t')\;.
\enq
The force ${\bf F}^{\rm PK(\nu)}$ is the
Peach-Koehler force from the external stress and it is given by
\beq
\label{Fdisloc}
F^{\rm PK(\nu)}_i({\bf r}_\nu)=-\epsilon_{ij}
             \sigma_{jk}({\bf r}_\nu)b^{(\nu)}_k\;.
\enq
Notice that in the case of interest here the external stress
is spatially inhomogeneous and the Peach-Koehler force 
on the $n$-th Burgers vector is evaluated at the location of
the Burgers vector. 

As we are interested in the dynamics of a neutral pair, it is convenient to
introduce the center of mass and relative coordinates of the pair as
\begin{eqnarray}
\label{CMcoord}
& & \bbox{\rho}={\bf r}_1-{\bf r}_2\;,\\\nonumber
& & {\bf R}_{CM}=\frac{{\bf r}_1+{\bf r}_2}{2}\;.
\end{eqnarray}
The Langevin equations for the center of mass and relative coordinates
of a single pair are then given by
\end{multicols}
\beq
{dR_{CM i}\over dt}= \frac{1}{2}\mu_{ij}
   \epsilon_{jk}b_l\Big[\sigma_{kl}\big({\bf R}_{CM}
                      +\bbox{\rho}/ 2\big)
       -\sigma_{kl}\big({\bf R}_{CM}-\bbox{\rho}/2\big)\Big]
      +\eta^{CM}_i(t)\;,
\enq
and 
\beq
{d\rho_i\over dt}= \mu_{ij}
   \epsilon_{jk}b_l\Big[\sigma_{kl}\big({\bf R}_{CM}
                      +\bbox{\rho}/2\big)
       +\sigma_{kl}\big({\bf R}_{CM}-\bbox{\rho}/2\big)\Big]
       -2\mu_{ij}\frac{\partial U_0(\rho)}{\partial\rho_j}
      +\eta_i(t)\;,
\enq
\begin{multicols}{2}
\noindent where $\bbox{\eta}=\bbox{\eta}^{(2)} - \bbox{\eta}^{(1)}$ and 
$\bbox{\eta}^{CM}=(\bbox{\eta}^{(2)} + \bbox{\eta}^{(1)})/2$.
Assuming that all dislocations have the same
mobility, i.e., $\mu^{(n)}_{ij}=\mu_{ij}$ for all $n$, we obtain 
\begin{eqnarray}
& &\langle \eta_{i} (t) \eta_{j}(t')\rangle
    =2k_BT\mu_{ij}\delta(t-t')\;,\\
& & \langle \eta^{CM}_{i} (t)\eta^{CM}_{j}(t')\rangle
      =\frac{k_B T}{2}\mu_{ij}\delta(t-t')\;.
\end{eqnarray}
We now assume that  both $R_1$ and the width of the disk, $R_2-R_1$, 
are much larger than $a_0$, so that the stress field can be considered uniform
on the scale $a_0$ of the initial separation between the dislocations 
of the pair.  We then expand the Peach-Koehler force about its value 
at the center of mass of the pair.
To leading order, we obtain
\begin{eqnarray}
& &{d{\bf R}_{CM}\over dt}\simeq\bbox{\eta}^{CM}(t)\;,\\
& &{d\rho_{i}\over dt}\simeq 2\mu_{ij}\bigg[
    -\frac{\partial U_0(\rho)}{\partial\rho_j}
     + \epsilon_{jk}\sigma_{kl}({\bf R}_{CM})b_l\bigg]
      +\eta_i(t)\;.
\end{eqnarray}
The center of mass of the pair
performs free thermal Brownian motion. We will focus below on the
dynamics of the relative coordinate.

The Langevin equation for the relative coordinate
can be simplified if we take into account that
climb motion is much slower than glide, yielding a 
strong anisotropy in the diffusion of dislocations. In general 
the mobility tensor can be written as
\beq
\label{mobility}
\mu_{ij}=\mu_{\rm glide} \hat{b}_i \hat{b}_j
    + \mu_{\rm climb}(\delta_{ij}-\hat{b}_i \hat{b}_j)\;,
\enq
where $\mu_{\rm glide}\;$ and $\mu_{\rm climb}$
are the mobility associated with the motion of a dislocation in 
its glide plane (defined by its Burgers vector and $\hat{\bf z}$) 
and perpendicular to the glide plane (climb), respectively,
and $\bbox{\hat{b}}={\bf b}/a_0$. Since climb can 
only occur with the creation of vacancies and interstitials, 
$\mu_{\rm climb}>>\mu_{\rm glide}\;$ \cite{Nabarro}. 
The motion of our edge dislocations is then predominantly
unidirectional, along the direction of the Burgers vector.
It is convenient to separate the relative displacement
of the pair, $\bbox{\rho}$, into its components along the glide and climb
directions, 
\beq
\label{rho}
\bbox{\rho}={\bf \hat{b}}x+({\bf \hat{z}}\times{\bf \hat{b}})y\;,
\enq
where $x={\bf \hat{b}}\cdot\bbox{\rho}$ and 
$y=({\bf \hat{z}}\times{\bf \hat{b}})\cdot\bbox{\rho}$.
The Langevin equation for the separation along the glide direction
is then given by
\beq
\label{langevin_1d}
{dx\over dt}\simeq 2\mu_{\rm glide} \bigg[
     -\frac{\partial U_0(\rho)}{\partial x}   
   +   a_0\sigma_\parallel({\bf R}_{CM})\bigg]
      +\eta_\parallel(t)\;.
\enq
where 
\beq
\sigma_\parallel({\bf R}_{CM})=\hat{b}_i\epsilon_{ij}\sigma_{jk}({\bf R}_{CM})\hat{b}_k\;,
\enq
and
\beq
\langle \eta_\parallel (t) \eta_\parallel(t')\rangle
    =2k_BT\mu_{\rm glide}\delta(t-t')\;.
\enq

The Langevin equation for the pair separation can also
be written in terms of the total potential energy of the neutral pair 
in the presence of
the external stress as
\beq
\frac{dx}{dt}=-2\mu_{\rm glide}\frac{\partial U(\rho)}{\partial x} 
     +\eta_\parallel(t)\;,
\enq
where
\beq
\label{int_total}
U(\rho) = U_0(\rho)+\frac{1}{2}\rho_i\big[
    \epsilon_{ij}\hat{b}_k+\epsilon_{ik}\hat{b}_j\big]
       \sigma_{jk}({\bf R}_{CM})t\;.
\enq
For the geometry of interest here the only non-vanishing component 
of the stress tensor $\sigma_{ij}$ is $\sigma_{r\phi}$ given in
Eq. (\ref{stress}). The neutral pairs that can be unbound by this stress 
when climb is forbidden are
those with Burgers vectors along the tangential direction of the disk.
Denoting simply by ${\bf r}$ the location of the center of mass of 
the pair relative to the center of the disk and assuming that the 
separation  $\sim a_0$ of the neutral pair is small compared 
to the width of the disk, the
interaction energy of the pair can be written as
\beq
\label{Udisloc}
U(\rho) \approx U_0(\rho)-a_0x\sigma(r)\;,
\enq
where  $\sigma(r)=-\sigma_{r\phi}(r)>0$ is obtained from  Eq.(\ref{stress}).


\begin{figure}
\begin{center}
\leavevmode
\hbox{%
\epsfxsize=3.in
\epsffile{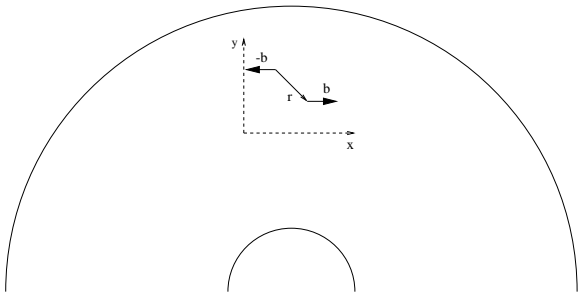}}
\end{center}
\caption{A pair of dislocations with opposite Burgers vectors. The Burgers
vectors are in the tangential direction.  The sketch shows the orientation 
of the pair, but it is
not to scale. Also shown is the local coordinate system with axis
along the glide and climb directions of the pair used in the text.}

\end{figure}


A plot of the interaction energy $U(x,y)$ 
as a function of the separation in the glide direction, $x$, 
for finite $y$ is shown in
Fig. 3. For simplicity we  neglect the angular part of
the interaction in zero shear and simply use Eq. (\ref{eff_int})
for $U_0$. This approximation does not change the qualitative
behavior of the results obtained below.
For a finite value of the applied stress, the interaction has
a saddle point at a location $x_0(y)$ 
on the positive $x$ axis 
(for $\sigma_{r\phi}<0$),
defined  by $[\partial U/\partial x]_{x=x_0}=0$ 
and given by the solution of
\beq
\label{saddle}
a_0\sigma\rho_0^2=x_0\overline{K}(\rho_0)k_BT\;,
\enq
where $\rho_0(y)=\sqrt{x_0^2(y)+y^2}$.
For small $y$, \cite{foot_angle}
\beq
x_0(y)\approx x_c-2\frac{y^2}{x_c}\;,
\enq
where $x_c$ is the solution of the equation
\beq
\label{saddlex}
a_0\sigma x_c=\overline{K}(x_c)k_BT\;.
\enq
At low temperatures, where the coupling constant can be
replaced by its bare value, we obtain 
\beq
\label{saddle_bare}
x_c(r)\approx \frac{K_0a_0}{4\pi\sigma(r)}\;.
\enq
The neutral pair can unbind by escaping over the barrier. 
This process creates 
a pair of free dislocations that then contributes to relaxing the applied 
stress. 


\begin{figure}
\psfrag{x}{$x/a_0$}
\psfrag{U}{$\frac{U4{\pi}}{K_0a_0^2}$}
\begin{center}
\leavevmode
\hbox{%
\epsfxsize=3.in
\epsffile{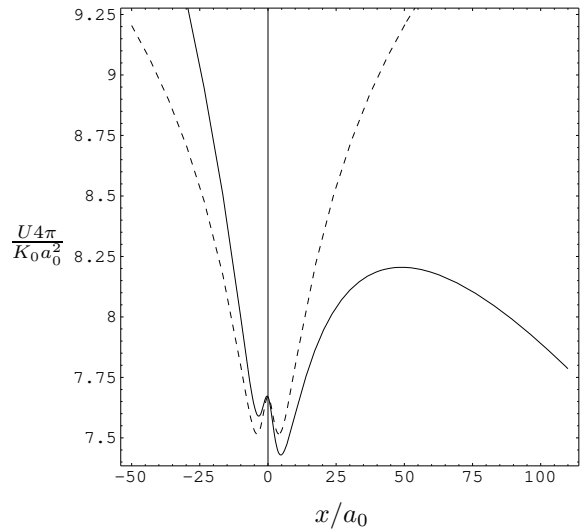}}
\end{center}
\caption{The interaction  energy  a neutral dislocation pair
in the presence of an applied shear stress as a function of 
the pair separation along the glide direction, $x$, for $y=4a_0$. 
The value used for the applied stress corresponds to $x_c=50a_0\;$. 
The dashed line shows the interaction energy of the pair for
$\sigma=0$. }
\end{figure}


At zero temperature the pair can be said to unbind when 
the saddle point $x_0$ occurs at a length scale
of order $a_0$. In thick samples, like the one used in the experiments
by the Argonne group,
this may indeed be the relevant mechanism for unbinding.
In thin films on the contrary, we will see that thermal activation plays
the dominant role.

To obtain a simple estimate of the critical value of the applied stress where
unbinding will occur at zero or very low
temperatures, we use the bare value of the pair 
binding energy and  consider a pair with $\theta=0$, i.e., $y=0$.
The saddle point is then given by Eq. (\ref{saddle_bare})
and unbinding can be estimated to occur when $x_c(r)\sim a_0$,
corresponding to a critical stress
\beq
\label{sigma0_cr}
\sigma_{cr}^0(r)\approx K_0/(4\pi t)\;.
\enq
This is the simplest ``criterion'' for the onset of plasticity. By solving
for $r$, one obtains the critical radius $r_{\rm pl}(I)$ where shear induced
dislocation unbinding occurs, yielding slippage of neighboring annular
sections of the vortex lattice,
\beq
\label{R_P}
r_{\rm pl}(I)=\bigg(\frac{I}{I_0-I}\bigg)^{1/2}
    \sqrt{\frac{2R_1^2R_2^2}{R_2^2-R_1^2}\ln\bigg(\frac{R_2}{R_1}\bigg)}\;,
\enq
where $I_0= cc_{66}t/B_{0}$ is the current where shear
induced proliferation of dislocations occurs across the entire sample.

If we
use typical parameters of the experiment by the Argonne group 
($R_1\approx 35 \mu m$, $R_2 \approx 350 \mu m$,
$t\approx 10 \mu m$, $H\approx 4 T$, and
$I\approx 15 mA\;$ in YBCO), we obtain
$r_{\rm pl}\approx 100\mu m$. The
critical radius $r_{\rm pl}(I)$ which marks the onset of plasticity increases with
current, indicating that, since the shear stress is largest near the inner
circumference of the disk, ``plastic flow'' occurs first in annular
sections close to the axis. This behavior is qualitatively consistent with
the experimental observations. An interesting result of our
simple model is that, at high fields, the current scale $I_0$, and therefore
$r_{\rm pl}(I)$, are {\it independent of the field}.

\section{Thermal unbinding}

At finite temperatures, neutral dislocation pairs can unbind by going over
the barrier via thermal activation. This is certainly the most relevant
mechanism in two dimensions or in thin samples.  The theory
of stress relaxation via nonlinear unbinding of dislocation pairs was
developed some time ago by Ambegaokar and collaborators \cite{AHNS80}
for superfluid films  and by
Bruinsma, Halperin, and Zippelius \cite{BHZ} for
two-dimensional crystals, and recently applied by Franosch and Nelson
\cite{franosch}
to describe nonlinear shear relaxation in smectic liquid crystal films.

Using standard methods, it is convenient to transform the Langevin equation 
for the pair separation $\bbox{\rho}$ into a
Fokker-Planck equation for the density of pairs,
$\Gamma(\bbox{\rho})$,  \cite{Chandra}, given by,
\beq
\label{FP}
\partial_t \Gamma=-\partial_i J_i\;,
\enq
where
\beq
\label{currentdisloc}
J_i=-\mu_{ij}\bigg[\bigg(\frac{\partial U_0}{\partial\rho_j}
                 -\epsilon_{jk}\sigma_{kl}({\bf r}) b_l\bigg)\Gamma
              +k_B T \partial_j\Gamma\bigg]\;
\enq
is the local current of pairs and ${\bf r}$ the center of mass of the pair.

Again, we separate the relative displacement
$\bbox{\rho}$ in components along the glide and climb
directions, as in Eq. (\ref{rho}),
and neglect  climb.
Introducing a diffusion constant for glide, $D=\mu_{\rm glide}/k_BT$,
and letting $u= U/k_BT$, the Fokker-Planck equation can be written as,
\beq
\label{FP1}
\partial_t\Gamma =-\partial_xJ_x\;,
\enq
with
\begin{eqnarray}
\label{FP2}
J_x(x,y) & = & -2D[(\partial_xu)\Gamma+\partial_x\Gamma]\\\nonumber
   & = & -2De^{-u}\partial_x(e^u\Gamma)\;.
\end{eqnarray}

In a steady state, Eq. (\ref{FP1}) reduces to $\partial_xJ_x=0$
and the current of dislocation pairs (which change their
separation in the glide ($x$) direction) is only 
a function of the pair separation in the climb ($y$) direction.
It is given by
\beq
\label{FPcur}
J_x(y)=2D~\frac{e^{u(x,y)}\Gamma(x,y)}{\int_{x}^\infty dx' e^{u(x',y)}}\;,
\enq
where $x$ is an arbitrary position (with $x>>a_0$, so that 
the shear stress can be considered
uniform).
For $x\rightarrow\infty$ we expect that the dislocation pair probability
density, $\Gamma$, vanishes rapidly, yielding
\beq
\label{FPcur2}
J_x(y)\int_x^\infty dx' e^{u(x',y})=2D~e^{u(x,y)}\Gamma(x,y)\;.
\enq
The integral on the left hand side of Eq. (\ref{FPcur2}) is dominated 
by the saddle point and the lower limit of integration can therefore be
extended to $-\infty$.  The integral is evaluated by the method of steepest descent,
with the result,
\beq
\int_{x}^\infty dx'  e^{u(x',y)}\approx 
        \sqrt{\frac{2\pi}{|[\partial^2_xu(\rho)]_{\rho=\rho_M(y)}|}}e^{u(\rho_M(y))}\;.
\enq

To evaluate $\Gamma(\rho)$ on the right hand side  of Eq. \ref{FPcur2}, 
we first notice that in the absence of external
stress the density $\Gamma$ will assume its equilibrium
form, $\Gamma_0(\rho)=(1/a_0^4)e^{-u_0(\rho)}\;$. 
For small stresses and $|x|<<x_M$, we  assume
that a local equilibrium ``barometric'' form  holds,
i.e.,  $\Gamma(x_2,y)=\Gamma(x_1,y)e^{-[u(x_2,y)-u(x_1,y)]}\;$. The
absence of climb implies that the integrated pair density at any $y$-cross
section remains the same before and after the application of stress,
\begin{eqnarray}
\int_{-\infty}^{\infty}dx_2\Gamma_0(x_2,y)
        & =&\int_{-\infty}^{\infty}dx_2 \Gamma(x_2,y)\\\nonumber
  & \approx &  \int_{-\infty}^{\infty}dx_2\Gamma(x_1,y)e^{-[u(x_2,y)-u(x_1,y)]}\;.
\end{eqnarray}
The main contribution to the integral in the last equality 
comes from the saddle-point of the full potential $u(x,y)$,
while the integral in the first term is dominated by the saddle point 
$x=0$ of the
potential in the absence of shear stress. After some algebra, we obtain
\beq
\Gamma(x, y)e^{u(x,y)}\approx \frac{1}{a_0^4}
        \bigg[\cosh\Big(\frac{a_0\sigma y}{k_BT}\Big)\bigg]^{-1}\;.
\enq
Finally, the current is given by,
\beq
\label{current_y}
J_x(y)\approx\frac{2D \sqrt{|[\partial^2_xu(\rho)]_{\rho=\rho_0(y)}|}}
     {\sqrt{2\pi}a_0^4\cosh\Big(\frac{a_0\sigma y}{k_BT}\Big)}
         e^{-u(\rho_0(y))}\;.
\enq
The main approximation used in obtaining this result is that of 
rare escapes over a 
high barrier. This requires  $\overline{K}>>1\;$, 
which is practically always satisfied as
$\overline{K}$ assumes its smallest value, 
$\overline{K}=4$, at $T_M$. We also must have $x_0>>a_0$, which imposes
an upper limit to the value of the  external stress.

The dissociation rate of bound dislocation pairs per
unit area of the lattice is given by
\beq
\label{dissocR}
R=\frac{d}{dt}\int_{-\infty}^{+\infty}dy\int_{x_0(y)}^{\infty}dx \Gamma(x,y)=\int_{-\infty}^{+\infty}dyJ_x(y)\;.
\enq
To evaluate this integral, we use a saddle-point approximation about
$y=0$, with the result,
\beq
\label{R}
R\simeq \frac{2D}{x_c^4}y^2(x_c)\frac{e^{\overline{K}(x_c)}}
          {\sqrt{\overline{K}(x_c)}}\;.
\enq
%
%
%

Dislocations of opposite Burgers vectors recombine when they come within
$x_0(y)\approx x_c$ of each other. The net production rate of free
dislocations is then given by \cite{AHNS80}
\beq
\label{netfree}
\frac{\partial n_f}{\partial t}=R-\langle v\rangle x_c n_f^2\;,
\enq
where $n_f$ is the areal density of free dislocations. 
The second term on the
right hand side of Eq. (\ref{netfree}) is the recombination
rate, with  $\langle v\rangle\approx a_0 \sigma D/k_BT$  the mean 
glide velocity of a free
dislocation under the shearing force $\sigma a_0\;$.
In the steady state the density of free dislocations is
given by $n_f\approx \sqrt{R/(\langle v\rangle x_c)}$, or
\beq
\label{n_D}
n_f(T,\sigma)\approx \sqrt{2}
     \frac{y(x_c)}{x_c^2}
    \frac{e^{\overline{K}(x_c)/2}}{\big[\overline{K}(x_c)\big]^{3/4}}\;.
\enq
For comparison, the  density of free dislocations in equilibrium, 
in the absence of external stress, is given by
\beq
n_{f}^0(T)=\frac{y_0}{a_0^2}=\frac{1}{a_0^2}e^{-E_ct/k_BT}\;.
\enq
The applied shear enhances exponentially $n_f(T,\sigma)$ 
over its equilibrium value. 
Equation (\ref{n_D}) is one of the main results of this section. 

To find the explicit dependence of $n_f$ on the applied stress,
we need to consider the relative importance of 
the two length scales entering the problem:
the location $x_c$ of the saddle point and 
the Kosterlitz-Thouless correlation length 
$\xi_-(T)$, which measures
the proximity to the melting transition below $T_M$.
The KT  correlation length $\xi_-$ is defined as the
length scale above which the scale-dependent interaction $\overline{K}(\rho)$
can be approximated by its
large distance limit,
$\overline{K}(\rho>>\xi_-)\approx\overline{K}(\rho=\infty)=4\big[1+\alpha(T)/4\big]$, with
$\alpha(T)\sim (T_M-T)^{1/2}$.
It is given by
\beq
\label{ximinus}
\xi_-(T)=a_0e^{1/\alpha(T)}\;,
\enq
and diverges as $T\rightarrow T_M^-$.
In this regime the fugacity is given by 
$y(\rho>>\xi_-)\approx \alpha(T)\rho^{-\alpha(T)/2}$.
If $x_c>>\xi_-$, the coupling constant and
the fugacity in Eq. (\ref{n_D}) can be replaced by
their large scale values and 
the density of free dislocations is given by
\beq
n_f\sim \overline \alpha(T)x_c^{-(2+\alpha(T)/2)}
                 \frac{e^{2(1+\alpha(T)/4)}}{(1+\alpha(T)/4)^{1/4}}\;,
\enq
which yields
\beq
n_f\sim \sigma^{2+\alpha(T)/2}\;.
\enq
The condition $x_c>>\xi_-$ or $4\pi\sigma/K_0<<e^{-1/\alpha(T)}$
will apply for sufficiently small stress $\sigma$, not too close to
$T_M$. As $T_M$ is approached from below, eventually one obtains
$\xi_->>x_c$. In this regime one can approximate \cite{AHNS80}
\begin{eqnarray}
& & \overline{K}(x_c)\approx 4\Big[1+\frac{1}{2\ln(x_c/a_0)}\Big]\;,\\\nonumber
& & y(x_c)\approx\big[\ln(x_c/a_0)\big]^{-1}\;.
\end{eqnarray}
The density of free dislocations is then given by
\beq
n_f\sim \frac{1}{x_c}^2\big[\ln(x_c/a_0)\big]^{-1}\sim \sigma^2\;.
\enq

For a Corbino disk of finite thickness $t$ the
corresponding  Kosterlitz-Thouless transition temperature
\cite{KostNelson} $T_{M}=c_{66}ta_0^2/4\pi k_B$ is very large,
so that all temperatures of interest are well below $T_M\;$.
In this region,
$\xi_-\sim a_0$ and the coupling constant and the fugacity 
in Eq. (\ref{n_D}) can
be replaced by their bare values. 
The density of free dislocations induced by the external stress
is given by
\beq
\label{nf_bare}
n_f(T,\sigma) \approx   n_{f}^0(T) \frac{a_0^2}{x_c^2}
    \frac{e^{\overline{K}_0/2}}{\overline{K}_0^{3/4}}
   ~\sim ~\sigma ^2\;.
\enq
%

At a finite temperature below $T_M$ we
can define a condition for ``shear-induced melting'' as the
value of the external stress where $x_c(\sigma)\simeq\xi_-(T)$.
Notice that for
$T<<T_M$, this condition reduces to the one used at $T=0$ 
to obtain Eq. (\ref{sigma0_cr}).
By solving this for $\sigma$ as a function of $T$,
we find that near $T_M$ the critical shear
stress $\sigma_c(r)$ where unbound dislocations proliferate in the vortex 
lattice is given by 
\beq
\label{sigmac_TM}
\sigma_c(r)\sim \frac{K_0}{4\pi}e^{-b/(T_M-T)^{1/2}}\;,
\enq
with $b>0$ a numerical constant.
The temperature dependence of this 
simple estimate is consistent with the experimental observation by the Argonne
group
that the applied stress (which is determined by the driving current) 
required for the
onset of plasticity decreases with increasing temperature (see Fig. 4 of
Ref. \onlinecite{Argonne}). 
%
%
By combining the various estimates for the critical shear stress where
unbound dislocations proliferate, i.e. the vortex lattice is 
``shear-melted'', we obtain the schematic phase diagram shown in Fig. 4.


\begin{figure}
\begin{center}
\leavevmode
\hbox{%
\epsfxsize=3.in
\epsffile{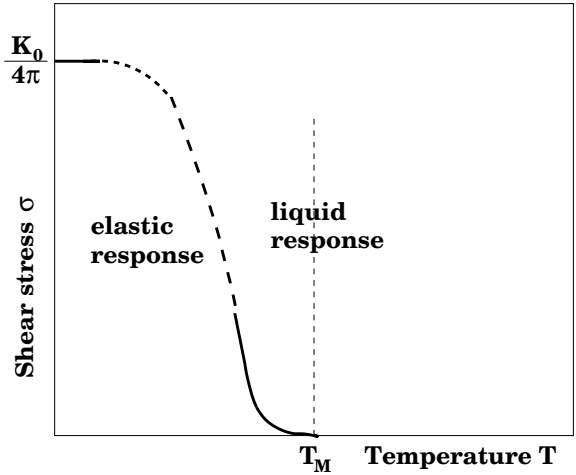}}
\end{center}
\caption{A schematic phase diagram in the $(\sigma,T)$ plane.
The dashed vertical line is the location of the Kosterlitz-Thouless
melting
temperature $T_M$ where thermal unbinding of neutral pairs occurs
in the absence of external shear. At a finite shear stress, 
free dislocations proliferate at the lower temperature denoted by
the thick continuous-dashed line. Even at $T=0$
a finite shear stress $\sim K_0/4\pi$ is sufficient to free bound pairs.
The dashed portion of the line separating the regions where neutral pairs are bound and
the vortex lattice responds elastically from the region where
free dislocations proliferate and the vortex arrays flows plastically
is an interpolation (by eye) between the low and high ($T\rightarrow T_M^-$)
estimates obtained in the text.}
\end{figure}


Finally, recalling that in the Corbino disk geometry the external
stress has the spatially inhomogeneous form given in Eq. (\ref{stress}),
we can solve Eq. (\ref{sigmac_TM}) for the critical radius 
$r_{\rm pl}(I,T)$ where shear-melting first
takes place for a given value of applied current and temperature.
We find that Eq. (\ref{R_P}) for the critical radius 
at $T=0$ generalizes simply to finite temperatures
as $r_{\rm pl}(I,T)$ is obtained from Eq. (\ref{R_P})
by the replacement
$I_0\rightarrow I_0(T)=I_0\exp\big[-b/(T_M-T)^{1/2}\big]$.

In the region below $T_M$ and above $\sigma_c(r)$, the driven vortex
array can be described as a lattice with a  concentration
$n_f(\sigma(r),T)$ of unbound dislocations, moving at a steady rate
along the direction of their Burgers vectors.
Moving dislocations
relax a strain. This provides a mechanism for nonlinear stress
relaxation that naturally yields nonlinear
IV characteristics in the superconductor. 

Assuming that the annular width of the disk is much larger than
$a_0$, so that a hydrodynamic description can be used,
we define a  Burgers-vector 
density,
\beq
\label{Burgers-density}
\bbox{\cal B}({\bf r},t)=\sum_\nu\bbox{\hat{b}}^{(\nu)}
   \delta({\bf r}-{\bf r}_\nu)\;,
\enq
which is related to the local strain in the lattice by
\beq
\epsilon_{ik}\partial_kw_{ij}=a_0{\cal B}_j\;,
\enq
where $w_{ij}$ is the unsymmetrized strain tensor. 
It is related to the  displacement field by $w_{ij}=\partial_iu_j$.  
In the presence of unbound
dislocations, the displacement field is no longer single-valued,
as indicated in Eq. (\ref{Burgers}), and
local deformations of the medium are more conveniently
described in terms of the local strain $w_{ij}$. 
Since free dislocations are always created in pairs, 
the Burgers-vector density is conserved,
\beq
\label{Bcons}
\partial_t{\cal B}_i+\partial_j{\cal J}_j^i=0\;,
\enq
where ${\cal J}_j^i$ is the current of the $i$-th component
of Burgers-vector density in the $j$-th direction,
namely
\beq
{\cal J}_j^i=\sum_\nu \hat{b}^{(\nu)}_i\frac{dr_{\nu j}}{dt}
  \delta({\bf r}-{\bf r}_\nu)\;.
\enq
The equation for the displacement field is replaced by an equation for 
the local strain, given by
\beq
\label{wij}
\partial_tw_{ij}=\frac{1}{n_0}\partial_i j_j+a_0\epsilon_{ik}{\cal J}_k^j\;,
\enq
where ${\bf j}= n_0{\bf v}$ (to linear order) is the number current density,
with ${\bf v}$ the local velocity of the vortex array.

In the Corbino disk geometry of interest here, the only nonvanishing component
of the dislocation current is in the azimuthal direction.
Combining Eq. (\ref{wij}) (specialized to a steady state) with
the results obtained in section IV, we estimate,
\beq
\partial_rv_\phi(r)=-a_0J_\phi^\phi\sim[\sigma(r)]^{1+\overline{K}(x_c)/2}\;,
\enq
where we have used Eq. (\ref{current_y}) evaluated at $y=0$ to obtain 
the dependence of the dislocation current on the shear stress.
The dependence on shear rate is highly nonlinear as $x_c\sim 1/\sigma(r)$.
The field from flux motion is radial and its magnitude is 
\beq
E(r)=\frac{1}{c} B_0v_\phi(r)\;.
\enq
The net voltage drop across the disk is given by
\beq
\Delta V=\int_{R_1}^{R_2} E(r) dr
\sim \int_{R_1}^{R_2} dr\int_{R_1}^r dr' [\sigma(r')]^{1+\overline{K}(x_c)/2}\;.
\enq
If $x_c>>\xi_-(T)$, we can replace $\overline{K}$ by its asymptotic value.
Since 
$\sigma(r)\sim I$, this immediately gives a nonlinear dependence
$\Delta V\sim I^{3+\alpha/8}$. The exponent $\alpha$ is nonuniversal,
with $\alpha\sim(T_M-T)^{1/2}$.

\section{Discussion}

We have presented a theoretical analysis of recent transport experiments 
on vortex lattices in the Corbino disk geometry. 
In the experiments, the onset of plasticity corresponds to the onset 
of nonlinearity in the IV characteristics and 
an unconventional scaling of the voltage with radial distance from the 
center of the disk.\cite{Argonne_Co1,Argonne,Argonne_Co2}
The nonlinear voltage-current scaling is found to be
$\Delta V\sim I^{3+\alpha(T)/8}$, with $\alpha$ a nonuniversal exponent
that depends on temperature.

In our theoretical model, the onset of plasticity is associated with 
the proliferation of free dislocations which break away from tightly 
bound pairs. In general, two mechanisms can be responsible for  
the nucleation of free dislocations from bound pairs: an externally applied 
shear stress and thermal fluctuations. 
In the absence of external shear, the thermally induced proliferation of 
free dislocations is simply the Kosterlitz-Thouless melting transition
of the lattice. A finite shear stress as the one imposed on the vortex lattice
by the external
current in the Corbino experiment can unbind dislocations 
below the KT transition temperature and even at $T=0$. 
Using standard methods of stochastic dynamics,  we have calculated 
the density of free dislocations as a function of  applied stress and 
the temperature. In contrast to earlier work on nonlinear shear relaxation
in superfluid,\cite{AHNS80} solid \cite{ZHN80,BHZ}
and smectic \cite{franosch} films, here the external stress is spatially
inhomogeneous. As a result, proliferation of unbound dislocations
occurs for different values of the external current at different locations
in the disk. 

The mechanism for the onset of plasticity proposed here,
namely the proliferation 
of free dislocations, is certainly relevant for thin Corbino disks, 
where the vortex lattice is essentially two-dimensional. 
In order to compare our results with the experiments we have assumed that 
even in a Corbino disk of finite thickness 
dislocations are essentially rigid over the thickness
of the sample {---} in other words screw dislocations are excluded.
In this case, the problem becomes essentially two-dimensional.
The energy cost for creating a dislocation pair is, however, proportional
to the sample thickness. This raises considerably the barrier
that bound pairs have to overcome to unbind via thermal activation.
The finite temperature calculation may therefore only be relevant 
in very thin samples, while in thick disks the physics is captured by the 
simple $T=0$ estimate described in Section 3.
Both at zero and finite temperature, our results, however, agree qualitatively
with the experimental observations, indicating that a simple picture of
stress-induced dislocation unbinding can account for the spatial dependence
of the onset of plasticity.
Specifically, we find that 
(1) ``shear-induced melting''  starts at the inner
boundary and propagates towards the outer rim of the disk as the driving 
current is increased, (2) the ``critical'' current for the onset of 
plastic slip of concentric vortex lattice planes decreases with 
increasing temperature, and (3) the calculated critical radius 
where plastic slip occurs for a fixed current
is of the same order of magnitude as the value measured in experiments.

A direction for future work is the study of the role of 
dislocation climb, which will of course require to couple dislocation
dynamics to that of
vacancies and interstitials. In addition, it would clearly be very interesting
to consider dislocation dynamics in the presence of quenched disorder.

\vspace{0.2in}
This work was supported by the National Science Foundation
through grant DMR-9805818.

\bibliography{panayotis}

\begin{thebibliography}{10}

\bibitem{CrabtreeNelson}
G.~W. Crabtree and D.~R. Nelson, Phys. Today {\bf {\bf 50}(4)},  38  (1997).

\bibitem{Blatter}
G. Blatter {\it et~al.}, Rev. Mod. Phys. {\bf {\bf 66}},  1125  (1994).

\bibitem{nels}
D.~R. Nelson,  in {\em Phenomenology and Applications of High-Temperature
  Superconductors} (Addison Wesley, Reading, Massachusetts, 1992), p.\ 187.

\bibitem{FisherFisherHuse}
D.~S. Fisher, M.~P.~A. Fisher, and D. Huse, Phys. Rev. B {\bf {\bf 43}},  130
  (1991).

\bibitem{NelsonVin}
D.~R. Nelson and V.~N. Vinokur, Phys. Rev. B {\bf {\bf 48}},  13 060  (1993).

\bibitem{mcmdrnhyd90}
M.~C. Marchetti and D.~R. Nelson, Phys. Rev. B {\bf {\bf 42}},  9938  (1990).

\bibitem{HuseMajum93}
D.~A. Huse and S.~N. Majumdar, Phys. Rev. Lett. {\bf {\bf 71}},  2473  (1993).

\bibitem{mcmdrn00}
M.~C. Marchetti and D.~R. Nelson, Physica C {\bf 330},  105  (2000) (e-print:
  arXiv:cond-mat/9909382).

\bibitem{Argonne_Co1}
D. L\'opez {\it et~al.}, Phys. Rev. Lett. {\bf {\bf 82}},  1277  (1999).

\bibitem{Argonne}
G.~W. Crabtree {\it et~al.}, J. Low Temp. Phys. {\bf {\bf 117}},  1313  (1999).

\bibitem{Argonne_Co2}
G.~W. Crabtree {\it et~al.}, Physica C {\bf {\bf 341-348}},  995  (2000).

\bibitem{mcmHouston}
M.~C. Marchetti, Physica C {\bf 341-348},  991  (2000) (e-print:
  arXiv:cond-mat/0007467).

\bibitem{AHNS80}
V. Ambegaokar, B.~I. Halperin, D.~R. Nelson, and E.~D. Siggia, Phys. Rev. B
  {\bf {\bf 21}},  1806  (1980).

\bibitem{BHZ}
R. Bruinsma, B.~I. Halperin, and A. Zippelius, Phys. Rev. B {\bf {\bf 25}},
  579  (1982).

\bibitem{ZHN80}
A. Zippelius, B.~I. Halperin, and D.~R. Nelson, Phys. Rev. B {\bf {\bf 22}},
  2514  (1980).

\bibitem{franosch}
T. Franosch and D.~R. Nelson, Phys. Rev. E {\bf {\bf 63}},  61706  (2001)
  (e-print: arXiv:cond-mat/0012433).

\bibitem{pertsinidis1}
A.Pertsinidis and X.~S. Ling, Phys. Rev. Lett. {\bf {\bf 87}},  98303  (2001)  (e-print: arXiv:cond-mat/0012306).

\bibitem{pertsinidis2}
A. Pertsinidis and X.~S. Ling, (e-print: arXiv:cond-mat/0103293).

\bibitem{KT_note}
See, e.g., D. R. Nelson, in {\it Phase Transitions and Critical Phenomena},
  Vol. 7, edited by C. Domb and J. Lebowitz (Academic, London, 1983) pp. 76-79.

\bibitem{paltiel00}
Y. Paltiel {\it et~al.}, Phys. Rev. Lett. {\bf {\bf 85}},  3712  (2000)
  (e-print: arXiv:cond-mat/0008092).

\bibitem{HallCorb}
F. Haenssler and L. Rinderer, Helv. Phys. Acta {\bf {\bf 40}},  659  (1967).

\bibitem{mcmdrnPhysica91}
M.~C. Marchetti and D.~R. Nelson, Physica C {\bf {\bf 174}},  40  (1991).

\bibitem{bc_note}
No-slip boundary conditions for the displacement yield a stress of the form
  given in Eq. (\ref{stress}), but with the plus sign inside the square bracket
  replaced by a minus sign. In this case the stress is largest (in magnitude)
  at the inner and outer rims of the disk and vanishes in the interior. Plastic
  slip will occur first at the rims, and then propagate towards the interior of
  the disk as the current is increased. No-slip boundary conditions are
  relevant for the experimental geometry proposed in Ref. \onlinecite{mcmdrn00}
  of a Corbino disk with Bose glass contacts.

\bibitem{mcmdis90}
M.~C. Marchetti and D.~R. Nelson, Phys. Rev. B {\bf {\bf 41}},  1910  (1990).

\bibitem{CarmenM}
M.~C. Miguel and M. Kardar, Phys. Rev. B {\bf {\bf 56}},  11 903  (1997)
  (e-print: arXiv:cond-mat/9706161).

\bibitem{nelson78}
D.~R. Nelson, Phys. Rev. B {\bf {\bf 18}},  2318  (1978).

\bibitem{y_note}
The same symbol $y$ is used here for the fugacity and below for the separation
  of a neutral pair of dislocations along the climb direction. There should be
  no confusion as he meaning of $y$ will be clear from the context.

\bibitem{peierls_note}
We have neglected the force arising from Peierls barriers to glide motion due
  to the periodicity of the lattice. In the Langevin equation for the
  separation of the dislocation pair in the glide direction considered below,
  this force only couples to gradients of the external stress.

\bibitem{Nabarro}
F.~R.~N. Nabarro and A.~T. Quintanilha,  in {\em Dislocations in Solids}
  (North-Holland Pub. Co., Amsterdam, 1980), Vol.~5.

\bibitem{foot_angle}
Taking into account the angular part of the dislocation interaction only
  changes the numerical coefficient of the $y^2$ term in the expression for the
  saddle point.

\bibitem{Chandra}
S. Chandrasekhar, Rev. Mod. Phys. {\bf {\bf 15}},  58  (1943).

\bibitem{KostNelson}
D.~R. Nelson and J.~M. Kosterlitz, Phys. Rev. Lett. {\bf {\bf 39}},  1201
  (1977).

\end{thebibliography}

\end{multicols}
\end{document}